\documentclass[english, aps, prfuids, amsmath, amssymb, preprint, longbibliography]{revtex4-1} 

\usepackage{color}
\usepackage[english]{babel}
\usepackage{graphicx}
\usepackage{verbatim}
\usepackage{ulem}
\usepackage[unicode=true, pdfusetitle, bookmarks=true, bookmarksnumbered=false, bookmarksopen=false, breaklinks=false, pdfborder={0 0 1}, backref=false, colorlinks=true]{hyperref}
\hypersetup{linkcolor=blue, citecolor=blue, urlcolor  = blue}

\graphicspath{ {./figures/} } 
\graphicspath{{obsolete/}}

\newcommand\eref\eqref

\newcommand\Reynolds{{\it Re}}
\newcommand\ManuscriptTitle{On the morphology of two-dimensional laminar vortex streets behind triangles}

\newcommand\citepar\cite
\newcommand\citebyname\cite


\begin{document}

\title{\ManuscriptTitle}
\author{Ildoo Kim}
\email{ildoo.kim.phys@gmail.com}
\affiliation{Department of Mechanical Engineering, Konkuk University, Chungju, South Korea 27478}
\date{\today}

\begin{abstract}
The two-dimensional laminar vortex streets behind a triangle have two morphologically distinct structures depending on the Reynolds number and the aspect ratio of the triangle.
These two structures are the conventional structure and the separated rows structure, where the latter is characterized by a thin layer of irrotational fluid between two vortex rows.
In this paper, by means of numerical simulation, we find that the separated rows structure occurs when the thickness of boundary layers is less than 25\% of their separation distance. 
We also show from the linear stability analysis that the criterion is related to the coupling of two boundary layers in producing unstable modes.
\end{abstract}

\maketitle

\section{Introduction}

The early investigation of vortex streets had focused on the interaction between vortices.
Assuming that the fluid is inviscid and incompressible and that the vortices have zero size, scientists investigated how the vortices are arranged into a certain spatiotemporal pattern.
Using this point vortex model, many asymptotic features were understood.
For example, von K\'{a}rm\'{a}n showed that the vortex arrangement should satisfy $h/\ell=0.28$, where $h$ is the transversal and $\ell$ is the longitudinal spacing between adjacent vortices \cite{vonKarman:1911vi}.
Another example is the Birkhoff's proposition that the Strouhal number, a non-dimensional frequency of vortex shedding, equals 0.2 when the Reynolds number is sufficiently large \cite{Birkhoff1953}.
In these classical studies, the vortices were assumed to exist in the fluid \textit{a priori}.
This was so because there is no physical mechanism to create the vorticity in the framework of inviscid theory.

In real fluid, however, it is impossible to ignore how vorticity is created and discharged into the fluid.
In general, the initial vorticity distribution has an impact on the downstream flow structures.
This is obvious when we consider the fact that most physical properties associated with the vortex shedding, such as the shedding frequency \citepar{Williamson:1998ti,Fey:1998vl,Jiang:2017by} or drag coefficient, \citepar{Roshko:1954vk,Henderson:1995vb,Chopra:2019dc} depend on the Reynolds number.
The vorticity is created by the action of viscosity in the boundary layers, and therefore it is important to investigate the effect of boundary layer on the downstream flow structure.

In this regard, we are interested in the vortex shedding behind triangles in two-dimension (2D).
Most studies of vortex streets concern the wake behind circles \citepar{Williamson:1996tn}, but they are not necessarily attractive to study the effect of boundary layers.
The diameter of a circle is a sole length scale in the system, and the resultant wake structure is also characterized by the a single parameter, that is, $\Reynolds$.
The use of circles is favored for simplicity, but it is limited in studying the effect of boundary layer because the thickness of a boundary layer scale with $1/\sqrt{\Reynolds}$ and thus cannot be controlled independently.
By using triangles, we have an additional control to vary the thickness of the boundary layers on two sides.
The triangles are characterized by two length scales, the base $D$ and the height $H$. 
Roughly, we posit that the thickness $\delta$ of boundary layers is proportional to the square-root of hypotenuse length, and by varying $H$ we control $\delta$ independent of $D$.
In addition, we limit the current investigation to 2D laminar vortex streets assuming the incompessibility.
Experimental studies show that the 2D vortex street is laminar up to $Re\sim1000$ \cite{Roushan:2005un}, and therefore the effect of boundary layer can be studied in a controlled way without undue complication due to three-dimensional characters of fluid.

The property of interest is the spatial arrangement of vortices.
Current study is motivated by a previous experimental study on vortex streets behind triangles \cite{Kim:2019kw}.
The paper reports that the vortex streets in 2D soap film channels can have two morphologically distinct spatial structures.
The first kind, the ``conventional mushroom'' (CM) structure, is like commonly observed vortex streets behind circles and is observed when $\Reynolds$ is relatively small.
The second kind, the ``separated rows'' (SR) structure, is characterized by the presence of a thin layer of irrotational fluid between two vortex rows and is observed when $\Reynolds$ is relatively high.
In SR structure, the irrotational fluid prevents the mixing of two vortex rows, delaying the breakdown of a vortex street.
This study \cite{Kim:2019kw} also suggests the criterion for SR structure that the thickness of boundary layer is to be thinner than a fraction of the base length of triangle.

The purpose of this work is to expand the previous study in two directions.
The first expansion is to reproduce the experimental observations in numerical simulations.
This is needed because the experiment was carried out in a flowing soap film channel \cite{Kellay:2017ie}, which is not always considered a Navier-Stokes system \cite{Chomaz:2001us, Kim:2017dn}.
In the literature, there are observations of a vortex structure similar to SR structure, with regards to the secondary instability behind circles, both experimentally \citepar{Cimbala1988,Vorobieff_2002,Wang_2010} and numerically \citepar{Inoue:1999wi,Kumar_2012,Dynnikova:2016is,Jiang:2019ea}.
However, the numerical simulation behind triangles is rare, except recent investigations using only equilateral triangles \cite{Mahato:2018bi,Zhu:2019hb}.
Therefore, the current study can build on credibility of the soap film experiments by simulating 2D vortex streets.
The second expansion is to provide a physical framework that allows us to understand the occurrence of SR structure.
Even though the previous study already proposed a criterion of SR structure, it was purely empirical and was not rationalized.
In this work, we rationalize the observation by modeling the base flow as a double shear layer.
The linear stability analysis implies that the SR structure occurs due to the decoupling of boundary layers on two sides of an object.

The paper is structured as follows: 
In section \ref{section2}, we report the methodology of numerical simulation.
In section \ref{section3}, the outcome of numerical simulation is discussed in a comparison to the experiments in literature. 
In section \ref{section4}, we propose a linear stability model that gives qualitative rationalization of the criterion of SR structure.
In section \ref{section5}, a brief summary is given.

\section{Method of numerical simulation}\label{section2}

The numerical simulation is carried out using a commercial software (COMSOL Multiphysics 5.3).
The software uses the finite element method to directly solve the Navier-Stokes equation,
\begin{equation}
\frac{\partial\vec{u}}{\partial{t}}+(\vec{u}\cdot\nabla)\vec{u}=-\frac{\nabla{p}}{\rho}+\nu \nabla^2\vec{u},
\label{eq:NSE}
\end{equation}
and the continuity equation,
\begin{equation}
\nabla\cdot\vec{u}=0,
\label{eq:cont}
\end{equation}
where $\vec{u}$ is the velocity field, $p$ is the pressure, $\rho$ is the density, and $\nu$ is the kinematic viscosity of fluid.
The commercial software provides a convenient tool for modeling the flow, meshing and the direct numerical simulation in a single package.

\begin{figure}
\begin{centering}
\includegraphics[width=5.6cm]{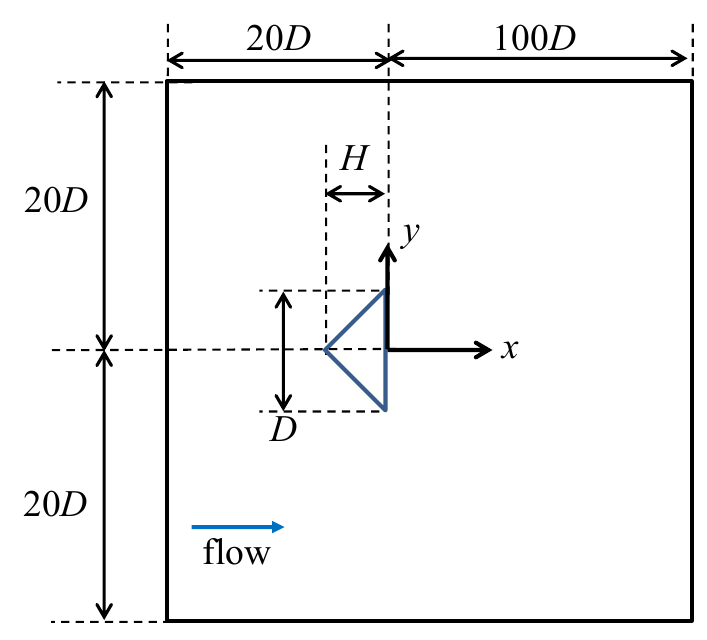}
\par
\end{centering}
\caption{
The flow model and computational domain.
A triangle of base $D$ and the height $H$ is placed at the origin.
The fluid flows in $\hat{x}$ direction, entering from the left and exiting to the right, at 65 cm/s.
The figure is not in scale.
\label{fig:experimentalsetup}}
\end{figure}

The flow geometry is modeled to simulate the vortex shedding in 2D as shown in a schematic diagram in Fig. \ref{fig:experimentalsetup}.
We first set the coordinate system such that $x$ is longitudinal and $y$ is transversal to the mean flow.
At the origin of the coordinate system, we place a triangle, whose base and height are $D$ and $H$, respectively.
These two length scales of triangles are independently varied; $D$ ranges from 0.02 cm to 0.1 cm, and $H'\equiv H/D$ ranges from 0.1 to 3.0.

The computational domain is large enough to cover the entire flow structure.
In the longitudinal direction, the domain covers from $20D$ upstream to $100D$ downstream, namely $x'\in[-20,100]$, where $x'=x/D$.
In the transversal direction, the domain spans $20D$ to the both direction, namely $y'\in[-20,20]$, where $y'=y/D$.
This translates to that our channel width is $40D$, which is wide enough to safely ignore any edge effect.
The flow enters the domain from the left boundary in Fig. \ref{fig:experimentalsetup} at a fixed speed $U=0.65\,\rm{m/s}$.
The properties of fluid are taken from the values of 2D soap film flows;
the density of fluid is $\rho=1\,\rm g/cm^3$, and the kinematic viscosity is $\nu=0.013\,\rm{cm^2/s}$ \cite{Kim:2019kw}.
Then the Reynolds number, defined as $\Reynolds=UD/\nu$, varies from 100 to 700.
In this range, the vortex streets in 2D are laminar \cite{Roushan:2005un}, therefore we do not need to consider the vortex mode transition that is known to occur near $Re\sim170$ in three-dimensional systems.

The above described flow model is solved by using the time-dependent laminar flow solver.
To reduce the computational cost, the shape and size of mesh are varied by location; near the triangular object, the size of quadrilateral meshes is as small as $ 5 \,\rm\mu{m}$, and away from the object, the size of triangular meshes is no bigger than $134\,\rm\mu{m}$.
For further information on the grid convergence, please see Appendix A.

Throughout the manuscript, the primed variables are used to denote dimensionless quantities normalized with respect to $D$, e.g. $H'=H/D$ and $\delta'=\delta/D$.

\section{Results}\label{section3}

\subsection{Qualitative observation of vortex structures}

\begin{figure*}
\begin{centering}
\includegraphics[viewport=0 40 764 530, width=16cm]{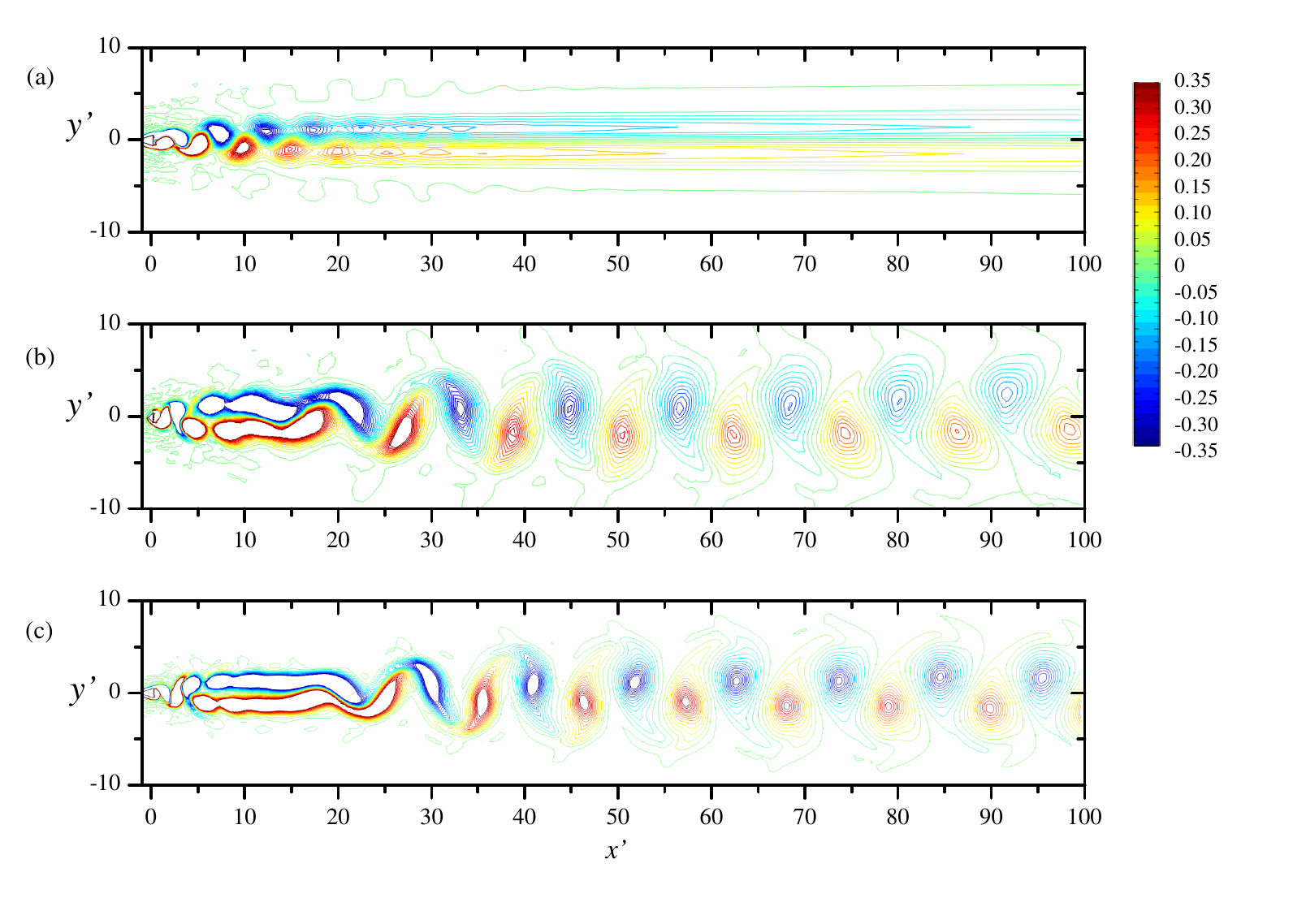}
\par
\end{centering}
\caption{
Snapshots of vorticity contours of simulated flows at 
(a) $H'=1.5$ and $\Reynolds=200$,
(b) $H'=0.5$ and $\Reynolds=200$, and
(c) $H'=1.5$ and $\Reynolds=300$.
The vorticity is normalized by $U/D$, and therefore the numbers associated with the color bar are the values of $(\nabla\times\vec{u})\cdot{D}/{U}$.
\label{fig:simulated_flow}}
\end{figure*}

In Fig. \ref{fig:simulated_flow}, the vorticity maps of three simulated flows are presented.
The images are calculated at (a) $H'=1.5$ and $\Reynolds=200$, (b) $H'=0.5$ and $\Reynolds=300$, and (c) $H'=1.5$ and $\Reynolds=300$.

The simulated flow in Fig. \ref{fig:simulated_flow}(a) resembles the commonly observed vortex streets behind circles and are labeled as the ``conventional mushroom'' (CM) structure.
The vortices detached from the triangle converges into an iconic vortex street pattern, where the longitudinal periodicity is approximately $4D$ \cite{Kim:2015jp}.
As vortices move downstream, their strength diminishes because of the mixing of two opposite signed vortices.
After some distance $\sim 30D$, the vortex street breaks down.

The simulated flows in Figs. \ref{fig:simulated_flow}(b) and (c) are morphologically distinctive from the one in Fig. \ref{fig:simulated_flow}(a).
The main character of these vortex streets is the thin layer of irrotational fluid separating two vortex rows, which spans from about $5D$ to $20D$, and we label these vortex streets as the ``separated rows'' (SR) structure.
This vortex structure is meta-stable; they undergo the secondary instability near $x=20D$, and the second vortex street has a longer wavelength $\sim10D$ compared to the same measure in CM structures.

We find that SR structure emerges in a very similar manner with the experimental report of the soap film flows.
First, for a given $\Reynolds$, SR structure emerges when $H'$ is smaller.
For example, the CM structure in Fig. \ref{fig:simulated_flow}(a) is simulated for $\Reynolds=200$ and $H'=1.5$.
While fixing $\Reynolds$ and decreasing $H'$ to 0.5, the vortex street becomes SR structure as seen in Fig. \ref{fig:simulated_flow}(b).
Second, for a given $H'$, SR structure emerges when $\Reynolds$ is greater. 
For example, Fig. \ref{fig:simulated_flow}(c) is simulated using $H'=1.5$, identical to (a), but by increasing $\Reynolds$ to 300, the structure became SR. 

\begin{figure}
\begin{centering}
\includegraphics[viewport=0 0 764 540, width=9cm]{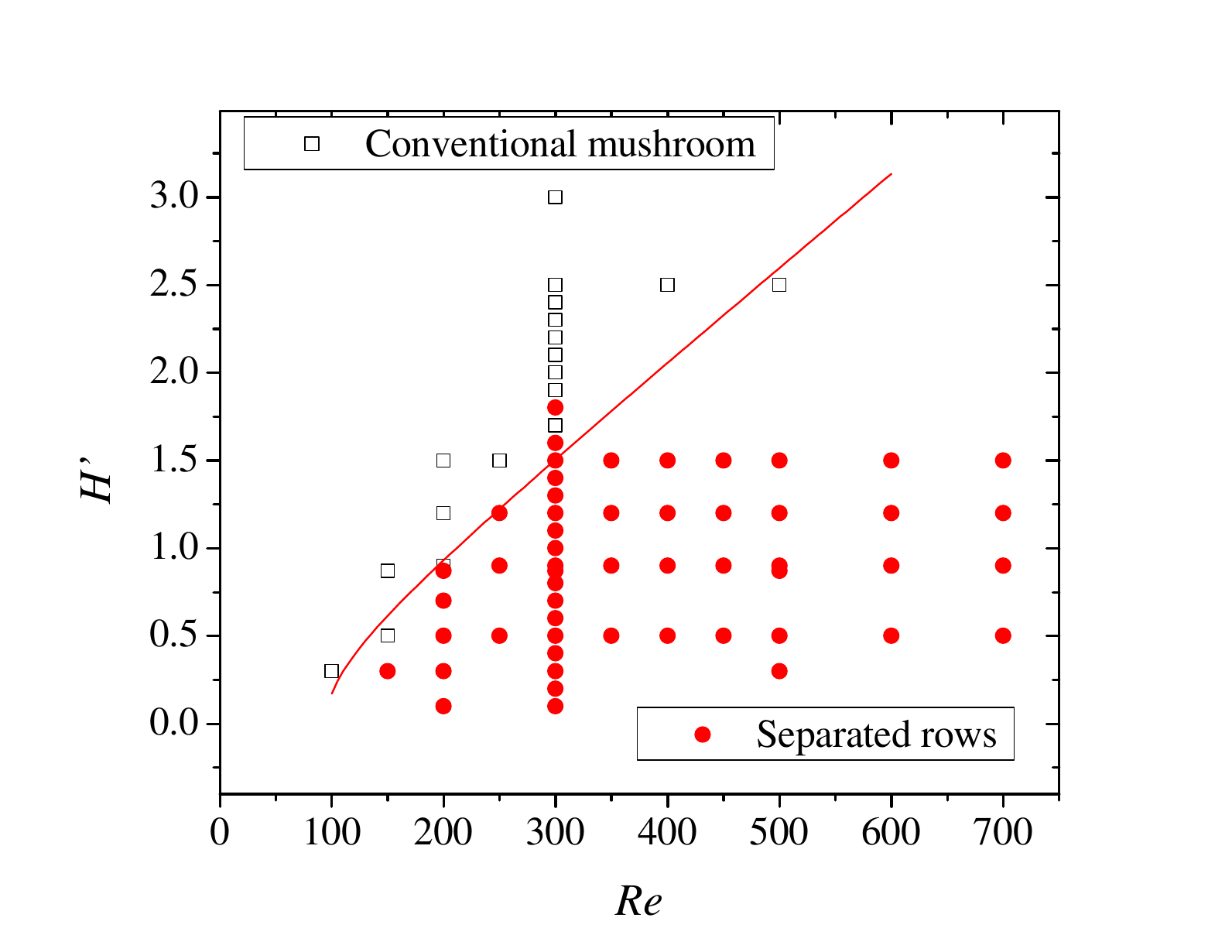}
\par
\end{centering}
\caption{
The map of the vortex structures behind triangles. 
When $\Reynolds<\Reynolds_{c2}$, the vortex street is conventional.
When $\Reynolds>\Reynolds_{c2}$, the exotic ``separated rows'' structure appears.
We find that $\Reynolds_{c2}$ is an increasing function with respect to $H'$.
The solid curve shows the calculation of \eref{eq:criterion}.
\label{fig:structure_map}}
\end{figure}

In Fig. \ref{fig:structure_map}, we summarized the morphology of vortex streets with respect to $H'$ and $\Reynolds$.
It is indicated that SR structure emerges at higher $\Reynolds$ when $H'$ is fixed. 
For example, when $H'=1.5$, the vortex street is in CM structure while $\Reynolds\leq250$, but as $\Reynolds$ goes beyond 300, the vortex street becomes SR structure.
We define a threshold Reynolds number $\Reynolds_{c2}$ such that the vortex street is in CM structure where $\Reynolds<\Reynolds_{c2}$ and in SR structure when $\Reynolds>\Reynolds_{c2}$.
Such defined $\Reynolds_{c2}$ is an increasing function of $H'$. 
For example, $\Reynolds_{c2}\approx200$ for the triangle of $H'=0.9$, but $\Reynolds_{c2}\approx300$ for the triangle of $H'=1.5$.

In short, our observation of simulated flows is consistent with the observation of real flows in a flowing soap film channel.
This indicates that the emergence of SR structure is not specific to a medium of experiment but a general phenomenon in 2D fluid.

\subsection{Condition for SR structure}

To establish a physical model to predict when vortex streets will have SR structure, we make an ansatz that the emergence of the separated rows structure is related to the ratio of the thickness of boundary layers $\delta$ and their separation distances $D$.
This proposition is based on and is consistent with the empirical proposition in the reference \cite{Kim:2019kw}.
Formally, we suggest that the vortex street is in SR structure if 
\begin{equation}
\delta' < \delta_0',
\label{eq:SR_condition}
\end{equation}
where $\delta'=\delta/D$ and the value of $\delta_0'$ to be determined later in this section.

\begin{figure}
\begin{centering}
\includegraphics[viewport=0 0 764 540, width=9cm]{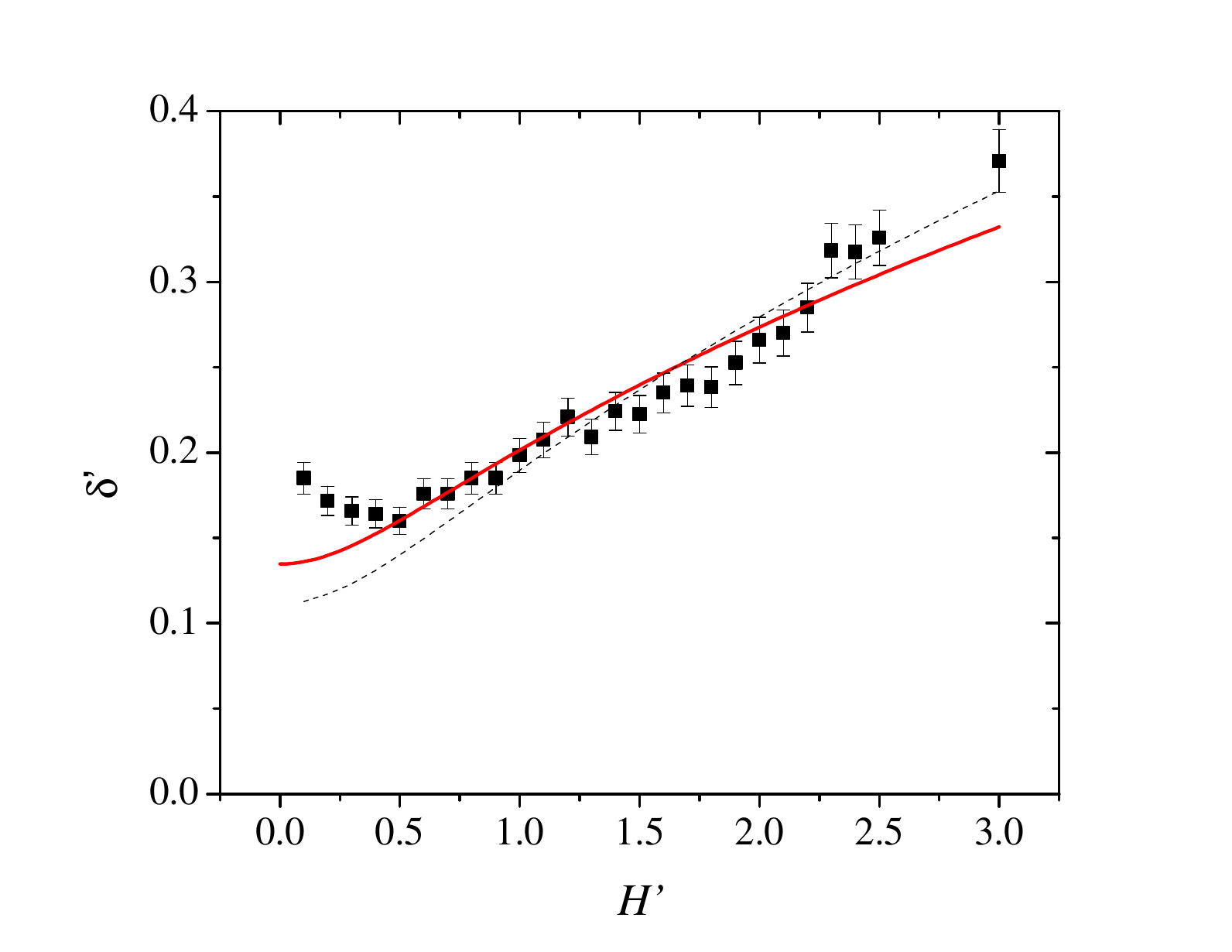}
\par
\end{centering}
\caption{
The boundary layer thickness $\delta'$ is plotted with respect to $H'$ at $\Reynolds=300$.
The solid and dash curves show the calculation of the Blasius boundary layer thickness in Eq. (\ref{eq:boundary_layer}) with $\alpha=3.3$ and the numerical solution of Falkner-Skan equation, respectively \cite{Schlichting}.
The error bars capture the frame-by-frame variations.
\label{fig:delta_ra}}
\end{figure}

The thickness of boundary layer, $\delta$, is the key control parameter of the ansatz, and its quantification is critical for the ensuing discussion.
In Fig. \ref{fig:delta_ra}, the evaluation of $\delta'$ from simulated flows is presented as a function of $H'$ for $\Reynolds=300$.
The evaluation is taken place at $x=0$ where the baseline of a triangle is located and is carried out by taking the width of the region where the vorticity has a non-zero value.
In general, the thickness of boundary layer is proportional to $\sqrt{\nu \Delta t}$, where $\nu$ is the fluid viscosity and $\Delta t$ is the time of interaction between fluid and the object.
We assume that the fluid passes by the length of triangle's hypotenuse $L$ at the free stream velocity $U$, so $\Delta t= L/U$.
Then it is inferred that 
\begin{equation}
\delta'=\alpha\frac{\sqrt{\nu\Delta t}}{D}=\alpha\frac{\sqrt{\nu (L/U)}}{D}=\alpha\sqrt{L'}\Reynolds^{-1/2},
\label{eq:boundary_layer}
\end{equation}
where $L'=L/D=\sqrt{H^2+(D/2)^2}/D =\sqrt{H'^2+0.25}$ and $\alpha$ is a proportionality coefficient.
By fitting the simulation data to the model in Eq. (\ref{eq:boundary_layer}), we find that the proportionality coefficient $\alpha\simeq 3.3$, as represented by the curve in Fig. \ref{fig:delta_ra}.

We note that in Fig. \ref{fig:delta_ra} there is a discrepancy between the evaluation of $\delta$ using the simulated flow and the model in Eq. \eref{eq:boundary_layer}.
This is due to the obstruction effect that was not taken into account by the model.
As the triangle becomes more obtuse, it slows down the flow significantly, and the approximation of $\Delta t=L/U$ is no longer valid.
Therefore, Eq. \eref{eq:boundary_layer} holds as a rough approximation.

The estimation of $\delta'$ can be improved by solving for the Falkner-Skan equation \cite{Schlichting}.
In Fig. \ref{fig:delta_ra}, the numerical calculation of the boundary layer thickness using the Falkner-Skan equation is plotted by a dash curve.
It agrees with the simulated flow a bit better than Eq. \eref{eq:boundary_layer}, but the improvement is not too substantial to abandon the convenience of using analytic expression of Eq. \eref{eq:boundary_layer}.

There is also a discrepancy between the estimation of $\alpha$ using the simulated flow and the real flow in literature \cite{Kim:2019kw}.
The value of proportionality constant, $\alpha\simeq3.3$, is smaller than the finding of experimental study using soap films ($\alpha\simeq 6$).
This may be due to the incorrect estimation of the viscosity of soap films.
Broadly speaking, the viscosity of a film is higher than the bulk liquid because the surfactants at the interface add extra molecular mobility \cite{Prasad:2009jj}.
The soap film experiment \cite{Kim:2019kw} did not take account the surface viscosity, and it is reasonable to consider that their viscosity is underestimated and that $\alpha$ is overestimated. 

Now, we derive the relationship between $\Reynolds_{c2}$ and $H'$ by using Eqs. (\ref{eq:SR_condition}) and (\ref{eq:boundary_layer}).
The substitution yields an inequality $\Reynolds>({\alpha^2}/{\delta_0'^2})L'$, and therefore
\begin{equation}
\Reynolds_{c2}=\frac{\alpha^2 L'}{\delta_0'^2}=\frac{\alpha^2}{\delta_0'^2}\sqrt{H'^2+0.25}.
\label{eq:criterion}
\end{equation}
Since we know $\alpha\simeq3.3$ as estimated from Fig. \ref{fig:delta_ra}, the only unknown parameter in Eq. (\ref{eq:criterion}) is $\delta_0'$.
This parameter can be estimated by data presented in Fig. \ref{fig:structure_map}, and we find that $\delta_0'\approx0.25$.
The curve in Fig. \ref{fig:structure_map} represents Eq. (\ref{eq:criterion}) with $\alpha=3.3$ and $\delta_0'=0.25$, and this line separates the CM regime and the SR regime well.
By putting the value of $\delta_0'$ Eq. \eref{eq:SR_condition}, we derive the condition for SR structure: 
\begin{equation}
\delta' <0.25.
\label{eq:SR_condition2}
\end{equation}

\section{Linear stability analysis}\label{section4}

\begin{figure}
\begin{centering}
\includegraphics[width=9cm]{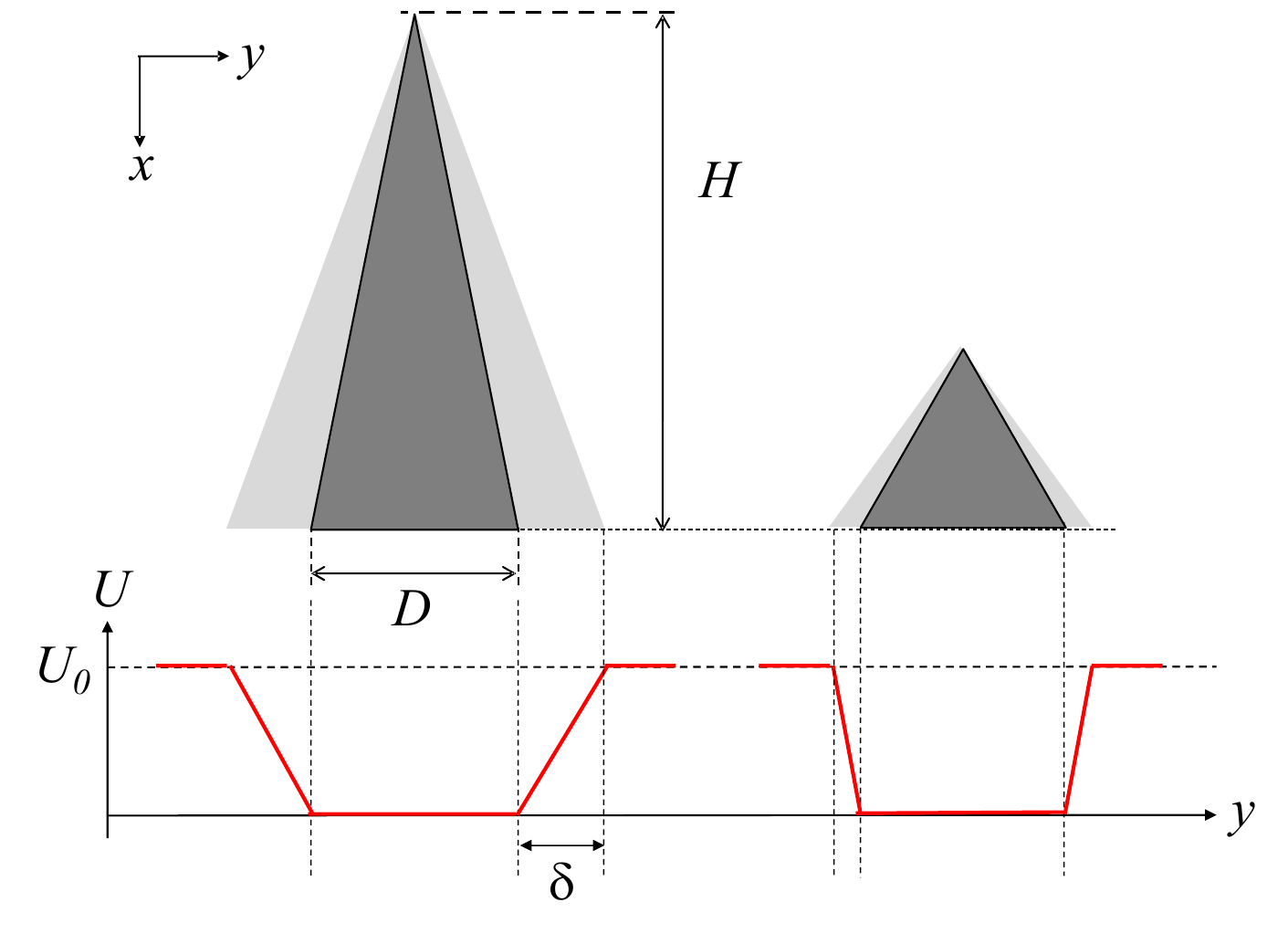}
\par
\end{centering}
\caption{
We model the base flow profile behind a triangle as a double shear layer.
For the same $D$, if $H$ is longer, the thicker boundary layer forms a gradual shear profile.
\label{fig:double_shear_layer}}
\end{figure}

To rationalize the criterion in Eq. (\ref{eq:SR_condition2}), we establish a physical model, in which we simplify the flow profile near the triangle as a pair of shear layers as depicted in Fig. \ref{fig:double_shear_layer}.
We concern the longitudinal component of velocity $U(y)$ at the baseline of triangle, i.e. $x=0$.
At the base of the triangle, we get $u=0$, and far from the triangle, we get $U=U_0$, where $U_0$ is the free stream velocity.
In between the two regimes, there are boundary layers of thickness $\delta$, and the velocity gradient of $U_0/\delta$, which also depends on $H'$, exists inside the boundary layer.

Formally, the base flow profile is expressed as
\begin{eqnarray}
U(y)&=&
\begin{cases}
U_0& \text{for }-\infty<y<-\frac{D}{2}-\delta\\
-\frac{U_0}{\delta}(y+\frac{D}{2}) & \text{for }-\frac{D}{2}-\delta<y<-\frac{D}{2}\\
0 & \text{for } -\frac{D}{2}<y<\frac{D}{2}\\
\frac{U_0}{\delta}(y-\frac{D}{2}) &\text{for }\frac{D}{2}<y<\frac{D}{2}+\delta\\
U_0 &\text{for }\frac{D}{2}+\delta<y<\infty.
\end{cases}
\label{eq:flowprofile}
\end{eqnarray}
Without loss of generality, we take $U_0=1$ and normalize $y$ with respect to $\delta$. 
In a dimensionless form, Eq. \eref{eq:flowprofile} becomes
\begin{eqnarray}
u(z)&=&
\begin{cases}
1& \text{for }-\infty<z<-\frac{1}{2\delta'}-1\\
-z-\frac{1}{2\delta'} & \text{for }-\frac{1}{2\delta'}-1<z<-\frac{1}{2\delta'}\\
0 & \text{for } -\frac{1}{2\delta'}<z<\frac{1}{2\delta'}\\
z-\frac{1}{2\delta'} &\text{for }\frac{1}{2\delta'}<z<\frac{1}{2\delta'}+1\\
1 &\text{for }\frac{1}{2\delta'}+1<z<\infty,
\end{cases}
\label{eq:flowprofile_dimensionless}
\end{eqnarray}
where $z=y/\delta$ and $\delta'=\delta/D$.
This model is an idealized form for the actual flow behind triangles, but it provides an acceptable approximation.
From the inspection of the simulated flow, it is found that the longitudinal component $u$ of the velocity roughly follows the model, and the transverse component $v$ is indifferent, having $|v/u|<0.16$.

To solve for the dispersion relation, we use Eq. \eref{eq:flowprofile_dimensionless} as an input for the Rayleigh's stability equation,
\begin{equation}
0
=(u-c)\left(\frac{\partial^2}{\partial{z}^2} -\alpha^2\right)\phi-\frac{d^2u}{dz^2}\phi,
\label{eq:stability}
\end{equation}
where $\alpha=k\delta$ is the non-dimensional wavenumber and $k$ is the dimensional wavenumber.
Here, $\phi$ is a part of the stream function of perturbation $u'$, i.e. $\vec{u}'=\nabla\times(\psi\hat{k})$ and $\psi=\phi(y)\exp[i\alpha(x-ct)]$.

Using the general solutions of Eq. \eref{eq:stability}, $\phi=\{ e^{{\alpha}z}, \,e^{-{\alpha}z} \}$, the trial solution is
\begin{eqnarray}
\phi(z)&=&
\begin{cases}
A_3e^{\alpha z}+B_3e^{-\alpha z} & \text{for } z<\frac{1}{2\delta'}\\
A_2e^{\alpha z}+B_2e^{-\alpha z} &\text{for }\frac{1}{2\delta'}<z<\frac{1}{2\delta'}+1\\
B_1e^{-\alpha z} &\text{for }z>\frac{1}{2\delta'}+1.
\end{cases}
\label{eq:trial_solution}
\end{eqnarray}
The five coefficients, $A_2$, $A_3$, $B_1$, $B_2$, and $B_3$ are to be determined by applying the boundary conditions.
The first boundary condition is that the $\phi$ is a symmetric function of $z$ so that the resultant oscillation is sinuous, which implies $A_3=B_3$.
The second boundary condition is that the pressure is continuous at the interfaces, i.e. $\Delta[(u-c)\phi'-u'\phi]=0$ across $z=1/2\delta'$ and $z=1/2\delta'+1$.
The third condition is that the normal velocity is continuous, i.e. $\Delta[\phi/(u-c)]=0$ across $z=1/2\delta'$ and $z=1/2\delta'+1$.

Applying the boundary conditions to the general solution in Eq. (\ref{eq:trial_solution}), the dispersion relation of small disturbance is derived in the following form:
\begin{equation}
c=\frac{-B\pm\sqrt{B^2-4AC}}{2A},
\label{eq:sol_c}
\end{equation}
where
\begin{eqnarray}
A&=&\alpha^2 e^\alpha (1+\tanh{\alpha/2\delta'}),
\\
B&=&
\tanh{\alpha/2\delta'}(\alpha^2e^\alpha-\alpha\sinh\alpha)+\alpha^2e^\alpha+\alpha\sinh\alpha,
\\
C&=&\alpha e^{\alpha} -\sinh{\alpha}.
\end{eqnarray}

Using the dispersion relation in Eq. \eref{eq:sol_c}, we calculate the growth rate of wavenumber $\alpha$.
Splitting $c$ into the real and imaginary parts $c=c_r+ic_i$, the stream function is written as $\psi\sim\exp[-i(\alpha c_r)t]\exp[\alpha c_i t]$.
It is, therefore, inferred that $B^2-4AC<0$ for an unstable mode exist.
The growth rate $\omega_d$ of a disturbance of the double shear layer is
\begin{equation}
\omega_d=\alpha c_i= \alpha\frac{\sqrt{4AC-B^2}}{2A}.
\label{eq:growth_double}
\end{equation}

The growth rate in Eq. \eref{eq:growth_double} can be compared to the growth rate of disturbance on a single shear layer.
The solution of the single shear layer $\omega_s$ is available in literature \citepar{Drazin-Reid}: $\omega_s=\frac{1}{2}\sqrt{\exp(-2\alpha)-(1-\alpha)^2}$ with the maximum growth rate of $\omega_s^m\simeq 0.2$ at $\alpha_m\approx0.79$.
We note that Eq. \eref{eq:growth_double} converges to the single shear layer case in the limit of $\delta'\rightarrow0$.

\begin{figure}
\begin{centering}
\includegraphics[viewport=0 0 764 550, width=9cm]{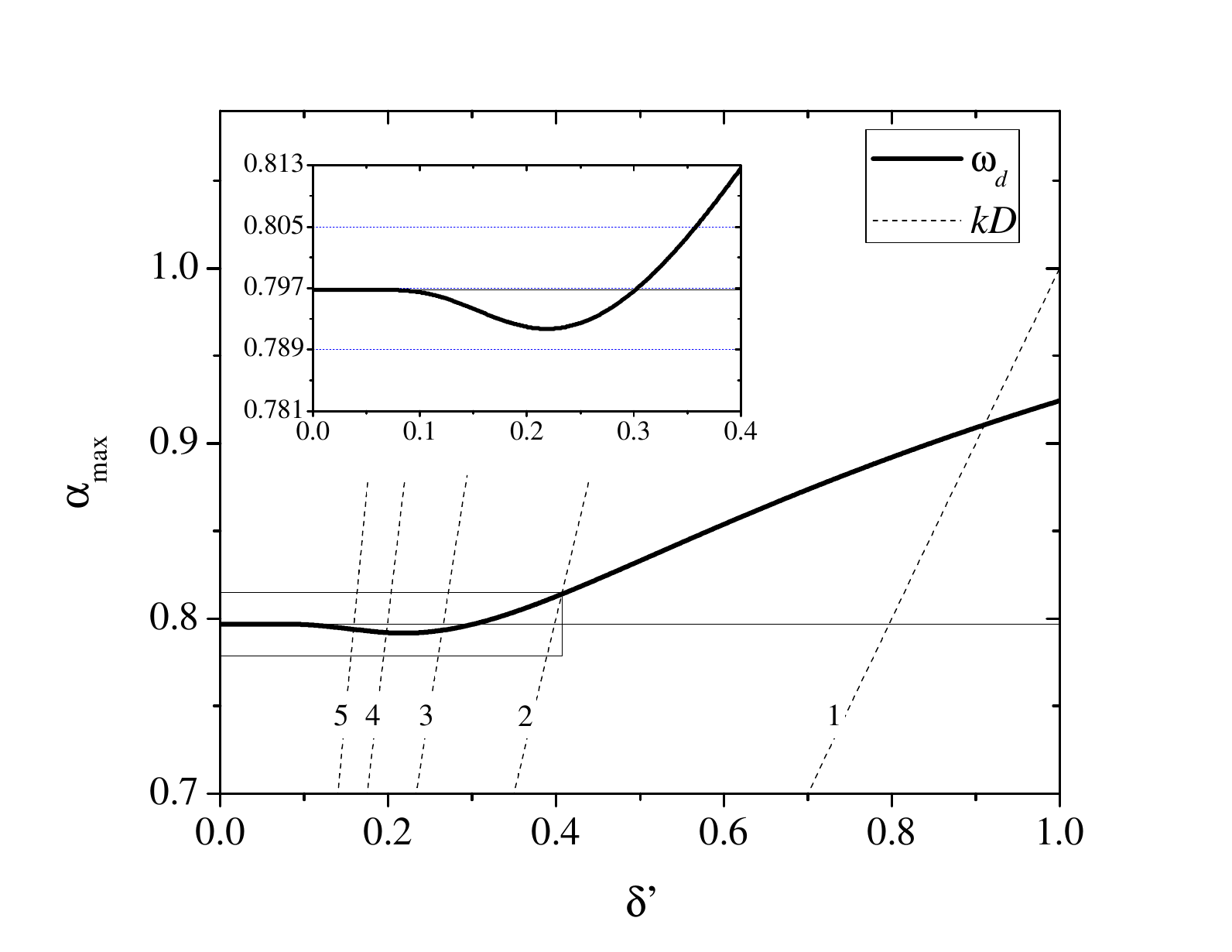}
\par
\end{centering}
\caption{
The fastest growing mode of the double shear layer profile in \eref{eq:flowprofile}.
When $\delta'\lesssim0.25$, $\alpha_{m}$ is approximately 0.8 and is similar to the single shear layer case.
However, when $\delta'>0.25$, the two cases are clearly distinctive.
The dotted lines represent $\alpha_{m}/\delta'=kD=$1, 2, 3, 4 and 5.
(inset) A close-up for $\delta'<0.4$.
\label{fig:dispersion_relation}}
\end{figure}

In Fig. \ref{fig:dispersion_relation}, we numerically calculate Eq. \eref{eq:growth_double} and plot the fastest growing wavenumber, $\alpha_{m}$ with respect to $\delta'$ from 0 to 1.
Interestingly, when $\delta'\gtrsim0.25$, $\alpha_{m}$ of the double shear layer and that of the single shear layer are significantly different.
However, when $\delta'<0.25$, the two models show subtle difference.
The result indicates that when $\delta'<0.25$, two shear layers are distant enough to be decoupled.
Due to the decoupling, the shear layers produce unstable modes separately. 
This contrasts to the case that when $\delta'>0.25$ the coupled oscillation produces the vortex shedding of shorter wavelength.

The linear stability analysis also hints at the question why SR structure is unstable or only meta-stable at best.
In Fig. \ref{fig:dispersion_relation}, five dotted lines are drawn to represent $\alpha/\delta'=kD=$1, 2, 3, 4, and 5, respectively.
As $\delta'$ decreases, the curve of the fastest growing mode intersects with the line of higher $kD$, indicating that the wavelength $\ell={2\pi}/{k}$ of a vortex street decreases.
Because the width $h$ of a vortex street is approximately proportional to $D$, i.e. $h\approx D$, the width-to-wavelength ratio is derived as $h/\ell\simeq kD/(2\pi)$.
When $\delta'=0.2$, $kD\simeq 4$, and the vortex street has $h/\ell\simeq0.64$.
However, such vortex arrangement is hydrodynamically unstable \citepar{vonKarman:1911vi}, and therefore SR structure should break down by itself after a finite lifetime.

\section{Summary}\label{section5}

We discussed the vortex streets behind triangles with the focus on their spatial structure.
By using the triangles instead of commonly used circles, the main feature of current work is to control the thickness of boundary layers and the Reynolds number independently.

The first part of investigation is to simulate the vortex streets using the computational method.
By inspecting the simulated flows, we found that the vortex streets can have two distinctive vortex structures; one being the conventional mushroom structure and the other being the separated rows structure.
In the latter structure, the vortex streets are characterized by a thin layer of irrotational fluid that separates two rows of opposite-signed vortices.
The observation of this exotic structure is consistent with and reinforces the same observation in a flowing soap film channel.
Further, our result suggests that the separated rows structure may occur in any other 2D fluid system.

The second part of investigation is to establish a simple model to answer why there are two distinctive structures.
We simplified the flow past a triangle as a double shear layer and performed the linear stability analysis.
The analysis shows that the two shear layers may or may not produce a couple oscillation depending on their separation distance.
Quantitatively, we found that the coupled oscillation occurs only when two shear layers are thicker than 25\% of their separation distance.
This observation is consistent with our observation of the simulated flows.

\bibliography{type2vortex2018}

\end{document}